\documentclass[aps,preprint]{revtex4-1}
\usepackage{graphicx}
\usepackage{textcomp}

\begin{document}

\title{Indistinguishable photons from deterministically integrated single quantum dots in heterogeneous GaAs/Si$_3$N$_4$ quantum photonic circuits}

\author{Peter Schnauber}
\affiliation{Institute of Solid State Physics, Technische Universit\"at Berlin, Berlin, Germany}
\author{Anshuman Singh}
\affiliation{National Institute of Standards and Technology, Gaithersburg, MD, USA}
\affiliation{Maryland NanoCenter, University of Maryland, College Park, USA}
\author{Johannes Schall}
\affiliation{Institute of Solid State Physics, Technische Universit\"at Berlin, Berlin, Germany}
\author{Suk In Park}
\affiliation{Center for Opto-Electronic Convergence Systems, Korea Institute of Science and Technology, Seoul, South Korea}
\author{Jin Dong Song}
\affiliation{Center for Opto-Electronic Convergence Systems, Korea Institute of Science and Technology, Seoul, South Korea}
\author{Sven Rodt}
\affiliation{Institute of Solid State Physics, Technische Universit\"at Berlin, Berlin, Germany}
\author{Kartik Srinivasan}
\affiliation{National Institute of Standards and Technology, Gaithersburg, MD, USA}
\affiliation{Joint Quantum Institute, NIST/University of Maryland, College Park, USA}
\author{Stephan Reitzenstein}
\affiliation{Institute of Solid State Physics, Technische Universit\"at Berlin, Berlin, Germany}
\author{Marcelo Davanco}
\email{marcelo.davanco@nist.gov}
\affiliation{National Institute of Standards and Technology, Gaithersburg, MD, USA}



\begin{abstract}
Silicon photonics enables scaling of quantum photonic systems by allowing the creation of extensive, low-loss, reconfigurable networks linking various functional on-chip elements. Inclusion of single quantum emitters onto photonic circuits, acting as on-demand sources of indistinguishable photons or single-photon nonlinearities, may enable large-scale chip-based quantum photonic circuits and networks. Towards this, we use low-temperature $\textit{in situ}$ electron-beam lithography to deterministically produce hybrid GaAs/Si$_3$N$_4$ photonic devices containing single InAs quantum dots precisely located inside nanophotonic structures, which act as efficient, Si$_3$N$_4$ waveguide-coupled on-chip, on-demand single-photon sources. The precise positioning afforded by our scalable fabrication method furthermore allows observation of post-selected indistinguishable photons. This indicates a promising path towards significant scaling of chip-based quantum photonics, enabled by large fluxes of indistinguishable single-photons produced on-demand, directly on-chip.
\end{abstract}

\maketitle

\section{Introduction}
In the development of advanced photonic quantum information systems, exemplified by various devised schemes for quantum simulation~\cite{sparrow_simulating_2018} and communication~\cite{orieux_recent_2016}, the ability to produce, manipulate and detect multiple identical photons in multiple spatial modes is a necessity. Integrated photonics has a great potential to fulfill such tasks, by allowing the creation of compact, complex, chip-scale photonic circuits that can implement phase-stable, reconfigurable, and integratable interferometric networks for linear optical operation at the single-photon level\cite{carolan_universal_2015,Harris2017}.

Silicon-based photonic integrated circuits are most promising for large system scaling, as foundry services offer the fabrication of user-designed, high quality integrated circuits comprising thousands of elements on shared wafer projects~\cite{Hochberg2010}. Importantly, photonic losses in on-chip waveguides and related linear elements - e.g., beam splitters and combiners, phase delay paths and linear filters - can be reduced sufficiently through design and process control, to enable significant scaling of integrated quantum photonic systems. Adding to a favorable set of characteristics, the introduction of solid-state quantum emitters~\cite{aharonovich_solid-state_2016} into silicon-based integrated quantum photonic circuits may yield unprecedented system scalability and functionality. Quantum emitters can e.g. act as high-rate, on-demand sources of indistinguishable single photons~\cite{Somaschi2016,Ding2016,Liu2019}, providing the large on-chip photon fluxes necessary for linear optical quantum systems such as boson sampling simulators~\cite{Wang2017,Loredo2017}. Emitters with optically addressable spins may furthermore act as stationary qubits in photonic networks, and, along similar lines, single-photon nonlinearities in single-emitter quantum cavity-electrodynamic systems~\cite{javadi_single-photon_2015,Sun2018} may allow  networks of deterministic quantum logic gates to be implemented.

In terms of silicon-compatible quantum light emitters, color centers in SiC have been shown to display promising optical and spin properties in a silicon-based material that is amenable to photonic integration~\cite{castelletto2014silicon,falk2013polytype}. Equally attractive emitters have not yet been identified in silicon or Si$_3$N$_4$. As a result, efforts to incorporate quantum emitters into photonic circuit platforms based on such materials have relied on hybrid integration with guest/host material systems that provide the desired optical properties. For instance, nitrogen-vacancy (NV) centers in diamond~\cite{mouradian_scalable_2015}, epitaxially grown InAs quantum dots (QDs) in GaAs~\cite{davanco_heterogeneous_2017}, and InAsP QDs in InP~\cite{zadeh_deterministic_2016} have been integrated with Si$_3$N$_4$ waveguides. In addition, InAs QDs in InP~\cite{Kim2017} on silicon-on-insulator, InAs QDs on GaAs-on-insulator~\cite{katsumi_transfer-printed_2018}, as well as carbon nanotubes~\cite{Khasminskaya2014} and 2D materials~\cite{sun2016optical} on silicon have also been shown. To date, however, Stranski-Krastanov (SK) self-assembled QDs have generally demonstrated superior optical coherence~\cite{Kuhlmann2015,Ding2016,Somaschi2016}, commonly evidenced by high degrees of two-photon interference, which is a pre-requisite to enable photon-photon interactions, e.g. in quantum gates. Thus, SK QDs currently offer the most favorable prospects for integrated quantum photonics.

One important drawback of the SK growth mode is the QDs' random spatial distribution across the growth surface. This imposes considerable challenges for maximizing light-matter interactions through nanophotonic geometries~\cite{lodahl_interfacing_2015}, which must be leveraged to create an efficient optical interface between the QD and the photonic circuit~\cite{davanco_heterogeneous_2017}. In such geometries, QDs must be positioned with high precision within the nanophotonic geometry, to maximize coupling to specific spatial modes, and at the same time the QD must be sufficiently far away from etched surfaces, to minimize effects detrimental to the QD coherence~\cite{Liu2018}. A number of methods have been developed for precisely locating individual SK QDs on a wafer surface, allowing subsequent fabrication of nanophotonic devices precisely located around selected dots~\cite{Dousse2008,Coles2016,unsleber2016highly,Sartison2017,Sapienza2015}. However, no hybrid devices have so far been demonstrated through such techniques~\cite{davanco_heterogeneous_2017,zadeh_deterministic_2016,Kim2017,katsumi_transfer-printed_2018}.

Here, we employ cryogenic cathodoluminescence (CL) spectroscopy and $\textit{in situ}$ electron beam lithography (EBL)~\cite{Gschrey2015b,Schnauber2018} to deterministically create hybrid integrated quantum photonic devices containing precisely positioned, pre-selected, individual InAs SK quantum dots. Our devices are based on a heterogeneous photonic integrated circuit platform, where GaAs devices containing positioned QDs are produced on top of Si$_3$N$_4$ waveguides~\cite{davanco_heterogeneous_2017}. We demonstrate triggered emission of single photons from a single QD in a hybrid nanowaveguide, coupled directly into a Si$_3$N$_4$ waveguide. In addition, we report the observation of two-photon interference, which indicates generation of post-selected indistinguishable photons from a single device. This is achieved through precise positioning of the single emitter at maximum distance from etched surfaces through our deterministic approach. Single-photon indistinguishability is essential for quantum photonic systems based on linear optical operations, and yet has never been reported in hybrid QD-silicon platforms. Our unprecedented results indicate good prospects for generation of on-demand indistinguishable photons in a scalable hybrid silicon photonic platform.

\section{Deterministic sample fabrication}

\begin{figure}[t!]
\centering
\includegraphics[width=14.6 cm]{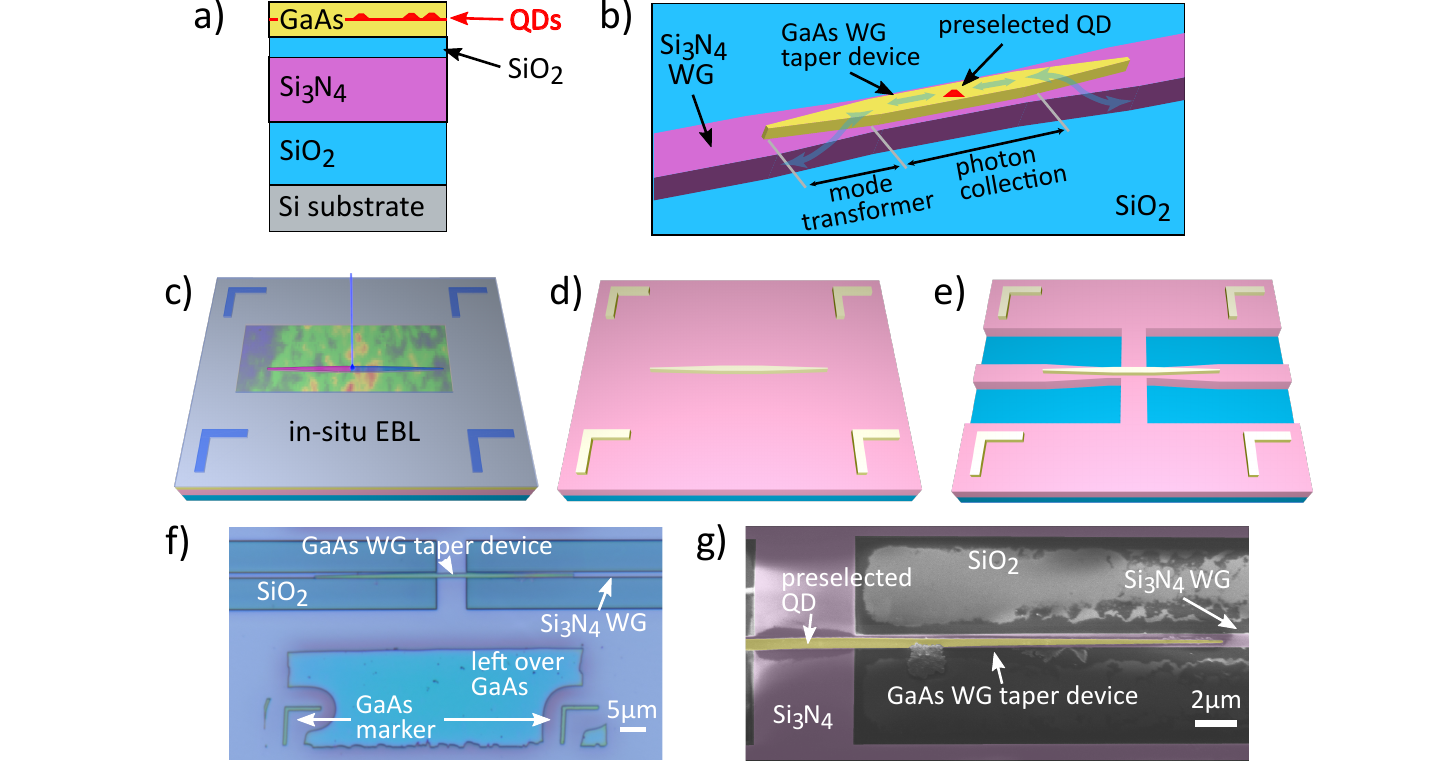}
\caption{a) Schematic layout of the wafer-bonded sample stack. b) Schematic GaAs-Si$_3$N$_4$ device design: The preselected InAs QD is hosted in a GaAs nanowaveguide that collects the QD's emission. The emission is then coupled into the Si$_3$N$_4$-SiO$_2$ WG using mode transformers. c)~-~e): Visualization of key sample fabrication steps: c) $\textit{in situ}$ EBL of a GaAs nanowaveguide pattern and markers aligned to a QD which was preselected using low temperature CL spectroscopy. d) GaAs nanowaveguide and markers on Si$_3$N$_4$ after etching the GaAs patterns and removing excess GaAs. e) fully-fabricated GaAs-Si$_3$N$_4$-SiO$_2$ WG device. f) false-color optical micrograph of fully-fabricated device QD~1. g) false-color SEM image of device QD~1, showing the GaAs WG taper (yellow) and the Si$_3$N$_4$ WG (pink).}
\label{fig:1fab}
\end{figure}

Efficiently interfacing individual QDs embedded in a III-V host with Si$_3$N$_4$ photonic waveguides has two requirements. First, the III-V host must be carefully shaped to support spatial modes into which emission from the individual QD can be efficiently funneled, and these modes must be simultaneously and efficiently coupled to Si$_3$N$_4$ WG modes~\cite{davanco_heterogeneous_2017}. Second, the QD must be located with high precision within the III-V host for optimal coupling to the desired spatial mode~\cite{lodahl_interfacing_2015}. In the hybrid devices of ref.~\cite{zadeh_deterministic_2016}, individual InAsP QDs were grown with high spatial precision within InP nanowires. Because such nanowires are generated through self-assembled growth, geometrical control of the QD-hosting InP is limited, which results in less efficient QD-waveguide interfaces - for instance, limiting the ability to create small mode-volume cavity modes for Purcell enhancement~\cite{lodahl_interfacing_2015}. Other groups~\cite{Kim2017,katsumi_transfer-printed_2018} have relied on lithography and etching to produce high-resolution, geometrically complex nanophotonic hosting geometries for embedded, randomly positioned SK QDs. No attempt has been made to position individual QDs precisely within the hosting geometries, however. In addition, in both demonstrations~\cite{Kim2017,katsumi_transfer-printed_2018}, QD-containing III-V devices were produced separately from the silicon photonic chip, then transferred onto the latter via pick-and-place processes, which offer limited scalability. The heterogeneous integration technique used in ref.~\cite{davanco_heterogeneous_2017} and this work, which starts from the wafer bonded stack in Fig.~\ref{fig:1fab}~a), allows for the creation of InAs SK QD-containing, complex GaAs nanophotonic devices directly integrated with Si$_3$N$_4$ waveguides. Here, this technique is combined with the cathodoluminescence spectroscopy and $\textit{in situ}$ electron-beam lithography of refs.~\cite{Gschrey2015b,Schnauber2018}, which has been shown to provide QD positioning accuracies of 34\,nm~\cite{Gschrey2015}. To realize tapered GaAs nanowaveguides for mode transformers, it is crucial to achieve device features sizes in the 50\,nm to 100\,nm range. Using the high patterning resolution of the $\textit{in situ}$ EBL along with proximity-correction grey-scale writing~\cite{Mrowinski2019}, feature sizes down to 50\,nm~\cite{Gschrey2015} can be reliably achieved. The combination of heterogeneous integration~\cite{davanco_heterogeneous_2017} with $\textit{in situ}$ EBL~\cite{Schnauber2018} therefore offers a deterministic, high resolution and scalable, purely top-down fabrication scheme. A comparison to other deterministic manufacturing approaches can be found in ref.~\cite{Schnauber2018}.

The $\textit{in situ}$ EBL technique has been used to produce a variety of photonic devices with deterministically positioned QDs, all on semiconducting (GaAs)~\cite{Gschrey2015b,Schnauber2018,Fischbach2017a,Mrowinski2019} or conducting (gold) substrates~\cite{Fischbach2017,fischbach2018deterministically}. For samples containing insulating layers like Si$_3$N$_4$ and SiO$_{2}$, as in this work, charging poses a major challenge. When the electron beam irradiates an insulating sample, the induced charge is not drained to the scanning electron microscope (SEM) ground and charges accumulate. This can already be a problem in standard EBL with positive tone resist doses on the order of 50$\frac{\text{\textmu}C}{\text{cm}^2}$, which require conductive polymers or thin metal films to be deposited onto the EBL resist. It becomes more severe for the $\textit{in situ}$ EBL which operates at electron doses in the 5000-50000\,$\frac{\text{\text{\textmu}C}}{\text{cm}^2}$ range. Small amounts of charging lead to electron beam deviations, and the fast build-up of large charge numbers leads to unstable beam jumps, which inhibit CL mapping or EBL patterning. High acceleration voltages reduce the amount of charge deposited in thin insulating layers~\cite{Yacobi1986}, but likewise the number of electron-hole-pairs created in the GaAs layer decreases and QD excitation becomes inefficient. The present work demonstrates that a sufficient balance can be achieved, allowing for high resolution QD positioning and pattern definition on heterogeneous substrates with thin insulating layers.

We fabricated a sample containing hybrid on-chip single-photon sources depicted schematically in Fig.~\ref{fig:1fab}~b), with varying geometrical parameters. Such sources are composed by a straight GaAs nanowaveguide section (labeled "photon collection") which hosts the preselected SK InAs QD and captures its emission into guided modes that are strongly confined in the GaAs ridge. Such GaAs-confined modes are subsequently converted into Si$_3$N$_4$ modes by adiabatic mode transformers implemented at the two ends of the photon capture section. As discussed in ref.~\cite{davanco_heterogeneous_2017}, such a geometry may offer QD-Si$_3$N$_4$ coupling efficiencies in excess of $90~\%$, through a combination of high photon capture probabilities and modal transformer efficiencies. In our samples, the photon collection regions consisted of 5\,\textmu m long straight GaAs ridges of widths >400\,nm, on top of a wide Si$_3$N$_4$ slab region of the same length, as seen in Figs.~\ref{fig:1fab}~e)~to~f). Over the mode transformer sections, the GaAs ridge was tapered from the central width down to 100\,nm at the tip, whereas the underlying Si$_3$N$_4$ waveguide maintained a width of $\approx 650$\,nm. The substrate cladding for the entire device consisted of thermal SiO$_2$. A 100\,nm thick spacer of SiO$_2$ is also featured between the GaAs and Si$_3$N$_4$ throughout the sample.

Fabrication started with a low-temperature plasma-bonded wafer stack consisting of a Si substrate, 3\,\textmu m thermal SiO$_2$, 250\,nm low-pressure chemical vapor deposition  (LPCVD) Si$_3$N$_4$, 100\,nm plasma-enhanced chemical vapor deposition (PECVD) SiO$_2$ and 190\,nm of GaAs containing SK InAs QDs at its center ~\cite{davanco_heterogeneous_2017}, as shown in Fig.~\ref{fig:1fab}~a). The plasma-bonded GaAs forms a uniform layer that spans over areas of several tens of square millimeters. GaAs nanowaveguides were deterministically patterned at the position of single preselected InAs QDs through $\textit{in situ}$ EBL~\cite{Schnauber2018,Gschrey2015b} as follows: The sample was coated with a dual-tone EBL resist which exhibits high contrast and high resolution at cryogenic temperatures~\cite{Kaganskiy2016}, mounted onto a custom-made liquid helium flow cryostat inside an SEM, and cooled to 7\,K. In the chamber, the sample was excited by an electron beam, and emitted light was collected through an NA = 0.8 elliptical mirror, then dispersed in a grating spectrometer. Through this process, spatially resolved CL spectrum maps over regions of hundreds square microns with 500\,nm steps were taken, while the electron dose remained well below the negative-tone onset-dose. Applying an acceleration voltage of 20\,kV and a beam current of 0.5\,nA, sample and resist charging was minimized while still operating at a high QD excitation. Comparing the CL imaging of sample regions with bonded GaAs to those without, we find that less charge accumulates in regions with GaAs, indicating beneficial charge carrier diffusion in the GaAs. On-the-fly spectral analysis of the CL maps yielded positions and spectra of suitable, individual QDs within a few minutes. Immediately after localization, proximity-corrected grey-scale $\textit{in situ}$ EBL~\cite{Mrowinski2019}, as illustrated in Fig.~\ref{fig:1fab}~c), was performed at 7\,K to define 400\,nm to 800\,nm wide and altogether 45\,\textmu m long symmetrical GaAs waveguide taper patterns (see Supplementary Material) aligned to the identified QDs with an uncertainty of about 55\,nm. The uncertainty is slightly higher than the previously achieved 34\,nm~\cite{Gschrey2015} due to a high QD density in the sample, which reduces the dynamic range of the 2D Gaussian fits used for QD localization if other emitters are spectrally and spatially within the fitting range. In addition, four L-shaped marker patterns, also aligned to the QD positions, were written outside the CL mapping area. The sample was then brought to room temperature, developed, and the resist pattern was transferred into the GaAs with an inductively coupled plasma reactive ion etch. Unpatterned GaAs was subsequently removed in a nitric acid/ceric ammonium nitrate aqueous solution, using a resist mask to protect the etched device and alignment mark areas, resulting in the intermediate sample layout shown in Fig.~\ref{fig:1fab}~d). The GaAs alignment marks (visible in the micrograph in Fig.~\ref{fig:1fab}~f)) were included for aligned EBL to be performed in a commercial 100~kV system. The Si$_3$N$_4$ waveguide patterns were defined as in ref.~\cite{davanco_heterogeneous_2017} and transferred via reactive-ion etching into the Si$_3$N$_4$ (Fig.~\ref{fig:1fab}~e)), and the sample was cleaved to allow endfire coupling to optical fibers inside of a cryostat.

Figure~\ref{fig:1fab}~f) shows a false-color optical micrograph of a finalized device. The positioned QD is located at the center of the 5\,\textmu m long, straight  portion of the GaAs nanowaveguide. Figure~\ref{fig:1fab}~g) shows a false-color SEM image of the same device, in which an unintended, vertical displacement of $\approx 60$~nm between the fabricated GaAs and Si$_3$N$_4$ waveguides is apparent. We note the GaAs marker dimensions and positions relative to the QD were manually calibrated in the $\textit{in situ}$ EBL system, which likely led to write field distortion and scaling errors. As a result, the markers featured imperfections that disallowed nanometer precision automatic alignment in the 100\,kV EBL system. Vertical displacements were systematically observed in all devices, and are likely due to the manual alignment procedure used for the fabrication of the Si$_3$N$_4$ layer, based on visual information from SEM scans and interferometric stage position readout. Incorporating the CL mapping system into state-of-the-art EBL equipment promises nanometer alignment fidelity throughout the whole process in the future.

Notably, some of the GaAs WG tapers that were fabricated with $\textit{in situ}$ EBL showed a bending at their left-hand side as shown in Fig.~\ref{fig:2det}~b)~and~c), while their right-hand side and the overall WG position remained unaffected. This bending stems from minor charging which occurs in parts of the sample and is explained in more detail in the Supplementary Material.

\section{Post-fabrication device characterization}

\begin{figure}[]
\centering
\includegraphics[width=14.6 cm]{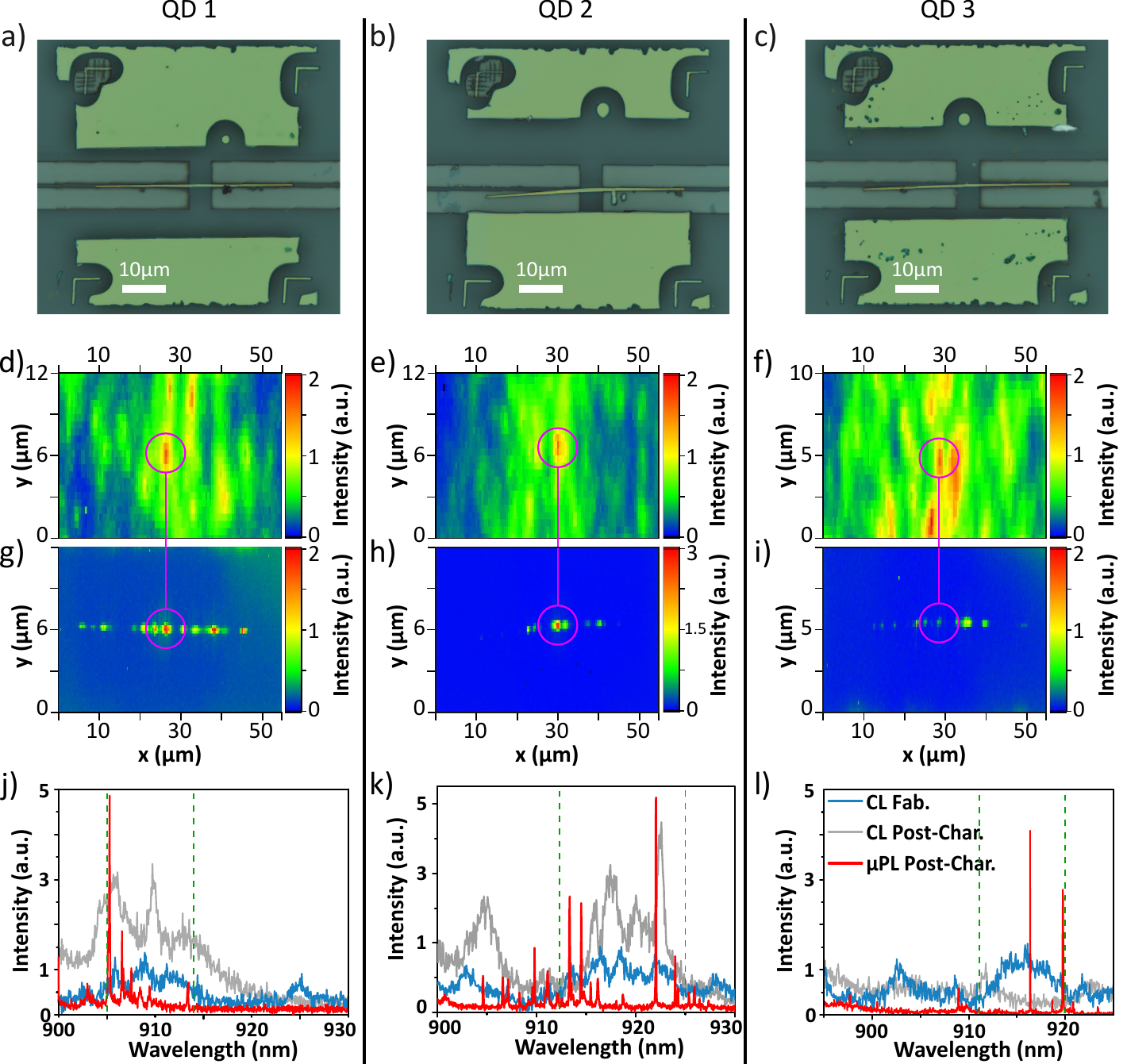}
\caption{a)~-~c): False-color optical micrographs showing devices QD 1-3, with GaAs colored in yellow for better contrast. The left-hand side of the GaAs WGs in b)~and~c) is bent downwards due to charging. d)~-~f): CL maps taken during $\textit{in situ}$ EBL to locate QDs 1-3. The pink circles mark the QD emission patterns that were used for QD localization. g)~-~i): CL maps taken on the fully-fabricated devices. Here, pink circles mark the GaAs WG center. The CL intensity in all maps d)~-~i) is integrated over those spectral regions, that were used to localize the QDs during the $\textit{in situ}$ EBL. These spectral regions are highlighted in j)~-~l) by green dashed lines. j~-~l): CL spectra taken during (after) fabrication in blue (grey) along with \textmu PL spectra after fabrication in red.}
\label{fig:2det}
\end{figure}

The successful fabrication of heterogeneous waveguide devices and the integration of preselected QDs was checked through conventional microscopy, scanning electron microscopy as well as micro-photoluminescence (\textmu PL) and CL spectroscopy on the fully fabricated sample. For \textmu PL, the sample was mounted inside a closed-cycle cryostat and cooled down to 7\,K. An off-resonant continuous wave (CW) laser at 821\,nm was focussed through an NA=0.28 microscope objective onto the designated QD position inside the GaAs WG. QD emission was collected from the Si$_3$N$_4$ WG endface through a lensed single mode fiber. CL maps and spectra are taken at a temperature of 7\,K, an electron beam current of 4\,nA and an acceleration voltage of 20\,kV.

Figure~\ref{fig:2det}~a)~-~c) show microscope images of three example WG devices QD~1, QD~2 and QD~3 with a GaAs nanowaveguide width of $\approx 620$\,nm that have been successfully positioned on preselected QDs. CL intensity maps of each QD taken during fabrication are visible in Fig.~\ref{fig:2det}~d)~-~f) and after fabrication in Fig.~\ref{fig:2det} g)~-~i). The maps show the CL intensity integrated over those spectral regions that were used to determine the QD positions during fabrication, marked by green dashed lines in Fig.~\ref{fig:2det}~j)~-~l). The pre- and post-fabrication CL maps for QD~1, QD~2 and QD~3 (Figs.~\ref{fig:2det}~d)~-~i)) display spatially matching, localized high intensity spots, marked by red pixels within the pink guide to the eye, that indicate successful, deterministic waveguide placement around the preselected individual QDs. Figure~\ref{fig:2det}~j)~-~l) show CL spectra with an exposure time of 50\,ms during (blue) and after (grey) fabrication from representative pixels at the positioned QD location, as indicated by each intensity maximum inside the pink guide to the eye in Fig.~\ref{fig:2det}~d)~-~i). \textmu PL spectra were obtained with an exposure of 1\,s after fabrication, but before the post-fabrication CL mapping, and are displayed in red. These spectra, which display broad CL, as well as sharp \textmu PL emission from individual QDs within the same spectral ranges, further support successful deterministic integration. The apparent discrepancy between the CL and \textmu PL spectra are due to differences in QD excitation conditions, where the various QD excitonic complexes are populated with different efficiencies. Also, in CL, the large injected charge density necessary for sufficient luminescence to be produced leads to spectrally broad QD lines. While QD~3 produced spectrally aligned pre-fabricated CL and \textmu PL emission, the post-fabrication CL emission displays considerably less intensity within the same spectral range. We believe that the repeated thermal cycles to which the sample was subjected, between the \textmu PL characterization and subsequent post-fabrication CL mapping, caused such degradation. Nonetheless, the pre-and post-fabrication CL maps still display the same localized intensity spots at the selected QD location. We note that all \textmu PL measurements were obtained by pumping the center region of the QD devices. Moving the pumping spot to other locations caused the emission spectra to vary considerably, as expected.

\section{Single-photon emission properties}
\begin{figure}[]
\centering
\includegraphics[width=14.6 cm]{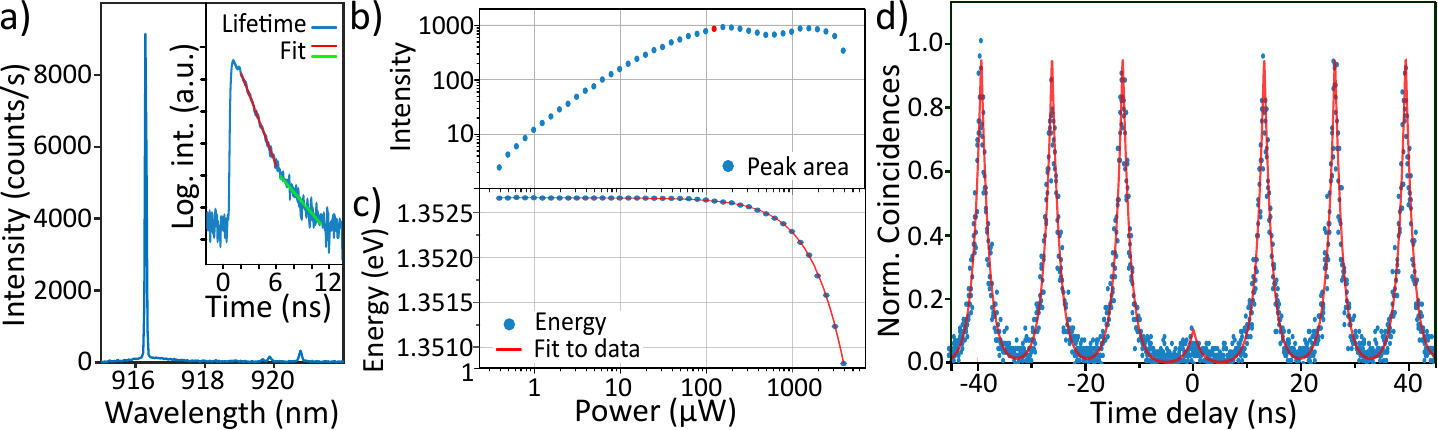}
\caption{a) PL spectrum of QD~3 in CW p-shell excitation. Inset: Natural logarithm of the excited state lifetime trace with two linear fits returning $\tau_{\text{r}}=1.39$\,ns and $\tau_{\text{r,2}}=3.15$\,ns. b)~PL intensity of the 916.3\,nm line over excitation power, in 1\,dB steps. c) Peak energy over excitation power (blue) and fit to data (red), showing a redshift with increasing excitation power. d) Pulsed excitation autocorrelation curve (blue) with fit to data (red) yielding $g^{(2)}(0)=0.11$.}
\label{fig:3spec}
\end{figure}

The finalized sample was tested inside of a closed-cycle cryostat at 7\,K. The sample was excited from the top using CW or pulsed tunable lasers through an NA=0.28 microscope objective, and the emission was captured from the Si$_3$N$_4$ WGs at the sample facet using aligned lensed single mode fibers inside the cryostat\cite{davanco_heterogeneous_2017}. Figure~\ref{fig:3spec}~a) shows a \textmu PL spectrum from the fabricated device housing QD~3, illuminated by a CW free-space laser beam tuned to $\approx904$\,nm, exciting the p-shell of the positioned QD~3 and giving rise to a narrow emission line at $\approx916.3$\,nm. The intensity of this PL line as a function of pump power is shown Fig.~\ref{fig:3spec}~b), showing saturation at a power of $155\,$\textmu W. The PL intensity was obtained as the peak area of a Gaussian model fit to each spectrum of the power series. Interestingly, a red-shift of the emission line is observed for increasing pump powers, see Fig.~\ref{fig:3spec}~c). This shift is likely due to a local increase in temperature in the GaAs WG, which lies on top of thermally insulating Si$_3$N$_4$-SiO$_{2}$. To investigate this hypothesis, we assume a linear temperature increase with excitation power due to linear absorption in the GaAs. With this model, we are able to faithfully fit the power-dependent spectral position of the QD line with a Bose-Einstein phonon law that describes the temperature dependence of the semiconductor bandgap~\cite{Ortner2004}, confirming a temperature-related effect. Moving towards strictly resonant excitation in the future, sample heating can be neglected, as the necessary pump powers are orders of magnitude lower than in p-shell excitation. More details and a comparison to temperature series measurements are given in the Supplementary Material.

We next measured the lifetime of the $\approx916.3$\,nm state by exciting QD~3 with a pulsed laser with a 76\,MHz picosecond pulse train at $\approx904$\,nm. The excitation laser was suppressed with a $\approx 500$\,pm free spectral range fiber-coupled grating filter with a transmission of $\approx60\,\%$ in addition to an edge pass filter. The filtered PL was detected on a superconducting nanowire single-photon detector (SNSPD) with a timing resolution of $\approx90$\,ps. The natural logarithm of the data is plotted in the inset of Fig.~\ref{fig:3spec}~a) and shows a double exponential decay. By fitting two linear curves to the natural logarithm of the data, we extract two decay constants of $\tau_{\text{r}}=(1.39\pm0.04)$\,ns and $\tau_{\text{r,2}}=(3.15\pm0.29)$\,ns (uncertainties are standard errors). The slower decay hints at a recapture process often seen in QDs~\cite{Peter2007}. To evaluate the single-photon emission purity of QD~3, pulsed excitation close to saturation was used, as indicated by the red dot in Fig.~\ref{fig:3spec}~b). The collected PL was split in a 50/50 fiber beam splitter and then detected by two SNSPDs (overall timing resolution $\approx 130$\,ps) in a Hanbury-Brown and Twiss (HBT) type configuration. Detection coincidences with time delay $\tau$ were tracked with a 64\,ps bin size. The normalized autocorrelation curve $g^{(2)}(\tau)$ is depicted in Fig.~\ref{fig:3spec}~d). The data was fitted with a two-sided exponential decay function convolved with a Gaussian that represented the experimental timing resolution, using a Poissonian statistics maximum likelihood estimator~\cite{Kirsanske2017a} (see Supplementary Material). Without any corrections, we obtain a conservative estimate of $g^{(2)}(0)=0.11\pm 0.04$ (uncertainty marks the 95\,\% confidence interval), clearly showing that the positioned QD~3 emits triggered single photons into the Si$_3$N$_4$ WG.

Next, we estimate the emission efficiency of our hybrid single photon source from the QD into the lensed fiber. During the HBT measurement, a combined photon stream of $\approx 50\,$kHz is measured on the detectors. Taking into account the grating filter and fiber transmission as well as the detector efficiency, we estimate a setup efficiency from collection fiber to detector of $\eta_{\text{Setup}} \approx 0.09$. Assuming 100\,\% quantum efficiency, the QD-to-fiber efficiency is $\eta_{\text{Source}} \approx 50\,\text{kHz} / (76\,\text{MHz}\cdot \eta_{\text{Setup}})\approx 0.7\,\%$ without further corrections. We compare this measured value with Finite Difference Time Domain (FDTD) simulations of a dipole radiating inside a geometry that closely approximates that of the real device, including fabrication imperfections. The simulations include the emission coupling from a dipole source into the 620\,nm wide GaAs WG, power transfer between the GaAs and Si$_3$N$_4$ guides, and coupling between the Si$_3$N$_4$ WG and collection lensed fiber (details in the Supplementary Material). Modelling the QD as linear dipole, we obtain $\eta_{\text{Source,1}} \approx 3\,\%$ and modelling the QD as rotating dipole resembling a trion state we find $\eta_{\text{Source,2}} \approx 1\,\%$. The linear dipole result gives a conservative upper bound for the optimum source efficiency $\eta_{\text{Source}}$ achievable with the WG examined here. We note that the simulated QD to GaAs WG coupling efficiency is $\approx 42\,\%$ for one propagation direction, and the overall coupling efficiency can potentially be significantly increased, by improving the fiber-to-Si$_3$N$_4$ waveguide coupling, the adiabatic mode transformer design, and introduction of a high reflectivity mirror on the back port of the GaAs waveguide~\cite{davanco_heterogeneous_2017}.

\section{Emission of indistinguishable photons in post-selection}

While the transfer of single-photon emission from QDs into silicon photonic circuits has been shown before using various sample preparation methods\cite{davanco_heterogeneous_2017,zadeh_deterministic_2016,Kim2017,katsumi_transfer-printed_2018}, the degree of indistinguishability of the emitted photons has never been evaluated in such hybrid systems. In fact, the close presence of dissimilar material interfaces to the QD introduces defect-rich regions that can reduce the QD coherence through electronic interaction with surface or defect states~\cite{Liu2018}, inhibiting emission of indistinguishable photons. A high degree of single-photon indistinguishability, however, is necessary for quantum photonic systems based on linear optical operations, and can serve as a baseline criterion for evaluating the quality of the fabrication process, regarding preservation of the QD coherence. In the following, we evaluate the coherence of photons emitted by the fully-fabricated device QD~3, pumped by CW laser light at $\approx904$~nm, close to saturation with an excitation power of 123\,\textmu W, marked by the red dot in Fig.~\ref{fig:3spec}~b).

\begin{figure}[]
\centering
\includegraphics[width=14.6 cm]{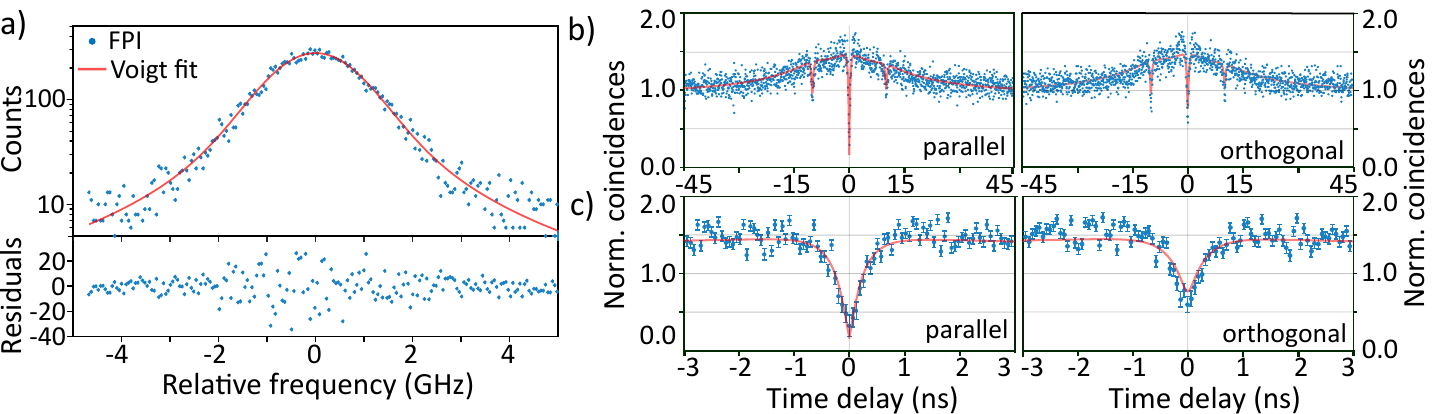}
\caption{a)~top: FPI spectrum of QD~3 (blue) with a fitted Voigt profile (red). Bottom: Voigt fit residuals. b) TPI coincidence curve (blue) and fitted model (red) for parallel (orthogonal) configuration is shown in the left (right) hand side panel. c) Same as b), including error bars and magnified around zero time delay for clarity.}
\label{fig:4coherence}
\end{figure}

As a first indicator of photon coherence, we measured the linewidth of the $\approx916.3$~nm emission line by passing it through a high resolution scanning Fabry-Perot interferometer (FPI) with a finesse of $\approx 200$ and free-spectral range of 40 GHz, and detecting the filtered signal with an SNSPD. The recorded spectrum is shown in Fig.~\ref{fig:4coherence}~a), where it is apparent that QD~3 has a linewidth of $\approx2$~GHz. Since there is no fine structure splitting the emission line stems most probably from a charged excitonic complex. Considering the GaAs nanowaveguide width of 620\,nm, the narrow linewidth suggests that severe degradation of the QD coherence through etched surfaces in the QD's vicinity~\cite{Liu2018} is avoided due to precise deterministic positioning. Fitting the FPI spectrum with a Voigt line function returns a more faithful result than Lorentzian or Gaussian shapes (see Supplementary Material) and we obtain a Voigt linewidth of $\Delta \Gamma_{\text{V}}=(2.20\pm 0.19)$\,GHz full-width at half-maximum (FWHM). A Lorentzian component of $\Delta \Gamma_{\text{L}}= (1.07\pm0.27)$\,GHz FWHM suggests homogeneous broadening beyond the Fourier limit of $\approx0.1$~GHz, likely due to dephasing from phonon interactions. The Gaussian component of $\Delta \Gamma_{\text{G}}= (1.54\pm 0.26)$\,GHz FWHM suggests inhomogeneous linewidth broadening due to spectral diffusion~\cite{Thoma2016}. All uncertainties mark the 95\,\% confidence interval. In our hybrid device both homogeneous and inhomogeneous broadening values are comparable to those observed under p-shell excitation in refs.~\cite{Thoma2016,Kirsanske2017a} and resonant excitation in ref.~\cite{nawrath2019coherence} in purely GaAs-AlGaAs-InAs-based samples. From the Lorentz linewidth we can extract an upper bound for the photon coherence time, yielding $\tau_{c,\text{FPI}} =1/ \pi \Delta \Gamma_{\text{L}} = (0.30\pm 0.03)$\,ns~\cite{Versteegh2014}.

Next, we measured the two-photon interference (TPI) contrast of subsequently emitted photons of the same $\approx916.3$\,nm line from QD~3, under the same CW excitation conditions, in a Hong-Ou-Mandel (HOM) type experiment. The fiber-collected PL was brought to a linear polarization state and coupled into a fiber-based unbalanced Mach-Zehner-interferometer with a $\delta \tau \approx10$\,ns arm imbalance. A variable half-wave plate in the long arm was used to align photons parallel or orthogonal to each other at the HOM beamsplitter and coincidences were measured with SNSPDs. The full HOM setup is detailed the Supplementary Material. The raw HOM autocorrelation traces for parallel and orthogonal photons are depicted in blue in Fig.~\ref{fig:4coherence}~b) and magnified around $\tau=0$ in Fig.~\ref{fig:4coherence}~c). The error bars in Fig.~\ref{fig:4coherence}~c) are the $1/ \sqrt{N}$ Poissonian uncertainty for each time bin with $N$ counts. At $\tau = 0$, the trace for parallel-polarized photons is clearly below 0.5 and below the orthogonal trace, marking the emission of indistinguishable photons. The bunching around $\tau=0$ hints blinking due to coupling of the QD to a dark state~\cite{nawrath2019coherence,Davanco2014}.

To extract an estimate for the coherence time $\tau_{c,\text{HOM}}$ and the two-photon interference visibility~$V$ we follow Ref.~\cite{Patel2008} and model the parallel and orthogonal coincidence traces with the functions $g^{(2)}_{\text{HOM},\parallel}(\tau)$ and $g^{(2)}_{\text{HOM},\perp}(\tau)$, respectively, where we include a HBT autocorrelation curve $g^{(2)}(\tau)$ with two separate two-sided exponentials to describe the bunching around zero time delay (see Supplementary Material for details on all functions). Fitting $g^{(2)}_{\text{HOM},\perp}(\tau)$ to the orthogonal case of Fig.~\ref{fig:4coherence}~b), we can extract $g^{(2)}(\tau)$. We use this to fit $g^{(2)}_{\text{HOM},\parallel}(\tau)$ to the parallel case of Fig.~\ref{fig:4coherence}~b), with $\tau_{c,\text{HOM}}$ and $V$ as the only free parameters. Both fits are plotted along the coincidence data in Fig.~\ref{fig:4coherence}~b) and~c). We obtain $\tau_{c,\text{HOM}}=(0.33\pm 0.12)$\,ns and $V=0.89^{+0.11}_{-0.29}$, where the uncertainty states the 95\,\% confidence interval. $\tau_{c,\text{HOM}}$ indicates a post-selection time window where indistinguishable photons are available. Since we are pumping close to saturation, our estimate for the coherence time $\tau_{c,\text{HOM}}$ represents a lower bound of what can be achieved in our system~\cite{Michler2002}. Within the uncertainty range, $\tau_{c,\text{HOM}}$ lies below the upper bound $\tau_{c,\text{FPI}}$ deduced from the FPI measurement.

Both the FPI spectrum analysis and the observation of two-photon interference indicate that our heterogeneous photonic integration platform can produce waveguide-coupled single-photon sources emitting light with a reasonable level of coherence. Both experiments were performed in quasi-resonant, and not strictly resonant excitation (a QD signal-to-pump laser noise ratio of about 1:2 was estimated for resonant excitation, which prevented observation of resonance fluorescence - see Supplementary Material for details). Since the QD was not excited resonantly - with which the highest optical coherence level can be achieved~\cite{Kuhlmann2015,reindl2019highly} - a clear-cut evaluation of the adversity imposed by our fabrication process upon QD coherence is not possible. In particular, QD~3 was not evaluated pre-fabrication, or even pre-wafer bonding, so its starting optical properties are unknown. Nonetheless, the QD linewidths and coherence times reported here are comparable with those observed from QDs in purely GaAs-based devices~\cite{Thoma2016,Kirsanske2017a,nawrath2019coherence} with which high degrees of two-photon interference were demonstrated, indicating good prospects for our technique.

\section{Discussion}
Precise alignment of the EBL patterns with respect to the QDs is essential to avoid excessive proximity to etched sidewalls, which may lead to degradation of quantum efficiency and, especially, coherence~\cite{Liu2018}. Our observation of a 2.2\,GHz linewidth from a positioned QD emission line, and subsequent demonstration of two-photon interference, indicate that the required precision can be met in our platform, and suggests that our fabrication method has minimal adverse effects on QD coherence. In order to increase the source efficiency $\eta_{\text{Source}}$ while preserving such high levels of photon coherence, photonic designs that avoid GaAs etched surfaces closer than 300\,nm to the QD~\cite{Thyrrestrup2018} while improving the emitter-WG-coupling are required. Creating sophisticated cavity-based devices~\cite{Hepp2018} for such a goal can be envisioned with our deterministic proximity-corrected grey-scale EBL process.

Even though CL excitation is restricted to conditions that avoid charging, which may be problematic for positioning precision and pattern resolution, we have been able to produce structures with widths $\leq$ 620\,nm with high precision around single, pre-selected QDs. We are able to consistently achieve 100\,nm thin waveguide taper tips through proximity-corrected grey-scale lithography in the $\textit{in situ}$ EBL step. In our samples, some level of charge draining is achieved through the GaAs top layer and Si substrate, so that, at a 20\,kV electron-beam acceleration voltage, charging due to the Si$_3$N$_4$-SiO2 layers is avoided, allowing sufficiently clear CL signals for QD positioning. Charging can nonetheless be further reduced through various measures: lower QD densities in the GaAs would require lower beam currents for efficient excitation and precise localization; a thinner bottom SiO$_{2}$ layer would absorb less energy without reducing optical confinement; electrically conductive polymers deposited on the resist would reduce charging in the latter and connect the GaAs to the silicon and the ground via the sample edge.

We note that multi-emitter experiments can be envisioned with our platform as well, with QDs that produce indistinguishable photons at identical wavelengths. While $\textit{in situ}$ EBL allows pre-selection of QDs with closely spectrally matching emission lines, with a <1\,nm accuracy, fine spectral tuning mechanisms are likely necessary. Local electrical control over the emission wavelength in GaAs p-i-n-doped layers through the quantum Stark effect is a promising approach, that is fully compatible with both the $\textit{in situ}$ EBL and heterogeneous sample stacks. This could also allow control of the QD charge environment, which has proven helpful in increasing the coherence of emitted photons~\cite{Somaschi2016}.

\section{Conclusion}
By applying $\textit{in situ}$ EBL to a heterogeneous GaAs~/ Si$_3$N$_4$ bonded wafer, we demonstrate the ability to deterministically produce GaAs nanophotonic devices with preselected and precisely located InAs QDs, which can be efficiently accessed by Si$_3$N$_4$ waveguides in an on-chip network. Our unprecedented demonstration of both triggered single-photon emission and post-selected indistinguishable photons produced by a single QD in a hybrid photonic circuit indicate a clear path towards highly scalable, chip-based quantum photonics. This can enable experiments such as Shor's algorithm~\cite{Politi2009} to be performed on-chip with triggered photons at rates substantially higher than currently available.

\section*{Funding}
A.S. acknowledges support by the Cooperative Research Agreement between the University of Maryland and NIST-CNST. The authors acknowledge funding from the German Research Foundation through CRC 787 'Semiconductor Nanophotonics: Materials, Models, Devices'.

\section*{Disclosures}
The authors declare that there are no conflicts of interest related to this article.
\bibliography{CLHQD_biblio}

\newpage
\section*{Supplementary Material: Indistinguishable photons from deterministically integrated single quantum dots in heterogeneous GaAs/Si$_3$N$_4$ quantum photonic circuits}

\section*{I. Charging effects during fabrication}
\setcounter{figure}{0} \renewcommand{\thefigure}{S\arabic{figure}} 

\begin{figure}[h]
\centering
\includegraphics[width=12 cm]{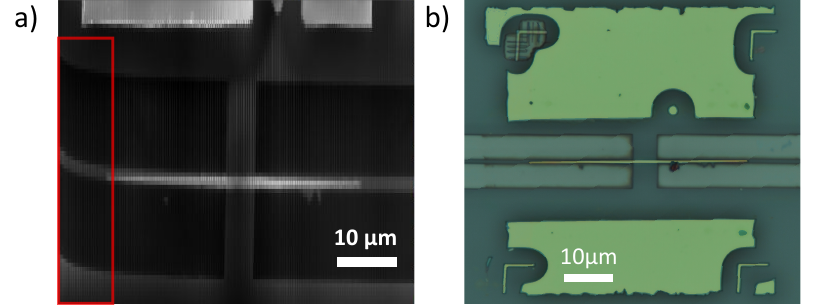}
\caption{a) SEM image of device QD~1 taken during post-fabrication CL mapping. In the first few seconds of the mapping process, the beam is deviated by a charge build-up, before stabilizing for the rest of the map. b) False-color optical microscope image of the same device proving that device QD~1 consists of straight WGs.}
\label{fig:S0charging}
\end{figure}

As can be seen from Fig.~2~b) and c) in the main text, some WGs show a downward-bent left-hand part. The WG bending amplitude correlates with the duration of the in-situ EBL process (stronger bending at later times during the 6\,h in-situ EBL run). It also correlates with the size of the GaAs layer area around the device (stronger bending for devices closer to GaAs layer borders). Both effects hint that during the in-situ EBL run, charge continuously accumulates and at some point cannot be dissipated sufficiently anymore. This leads to a charge build-up in the mapping process that deviates the beam in the beginning of the patterning step. This effect can be reproduced in the post-fabrication mapping: Because no planar GaAs exists around the fabricated devices anymore, there is insufficient charge dissipation and the beam is deviated in the beginning of the post-fabrication mapping process, as can be seen in the simultaneously acquired SEM image in Fig.~\ref{fig:S0charging}~a). Fig.~\ref{fig:S0charging}~b) depicts an optical micrograph of the same WG device, showing that it is not bent. The beam shift is analogous to the deviations for devices fabricated late in the in-situ EBL run. In the post-fabrication maps Fig.~2~g)~-~i) in the main text, beam deviations at the GaAs devices were avoided by starting the map far enough away for the beam to stabilize before scanning the GaAs devices. As discussed in the main text, lower QD densities requiring lower beam currents, as well as a thinner SiO$_2$ layer and conductive polymers on top of the EBL resist are expected to significantly reduce charging in the future.

\newpage
\section*{II. GaAs waveguide taper pattern}
\begin{figure}[h]
\centering
\includegraphics[width=14.2 cm]{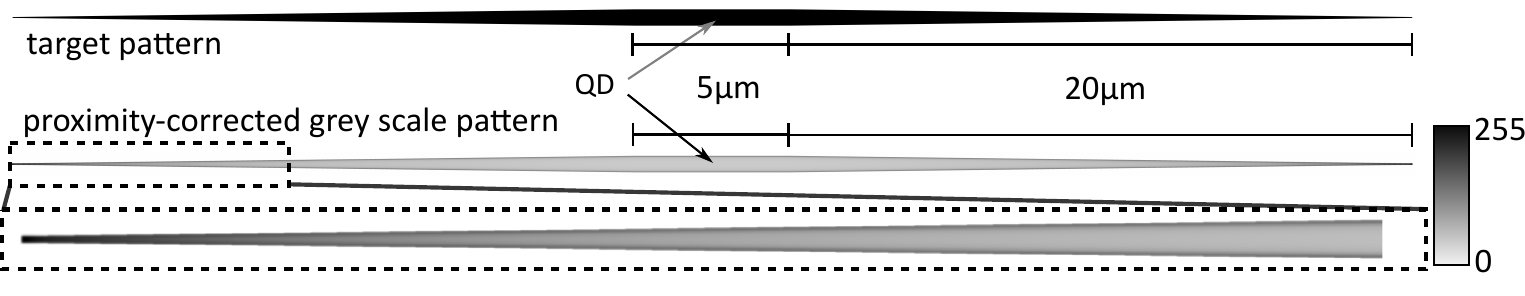}
\caption{top: Target GaAs nanowaveguide pattern used to integrate QD~1-3. btm: Proximity-corrected grey-scale EBL pattern for in-situ EBL to manufacture the target pattern along with a close-up of the grey-scale taper-tip for better visualization of the grey scales.}
\label{fig:S01pattern}
\end{figure}

The GaAs nanowaveguides, which host the InAs QD, are patterned using proximity-corrected grey-scale EBL. This is particularly important as the electron dose per pixel needed for cross-linking the resist increases by a factor of more than 4 at the taper tip, as compared to the taper center. Fig.~\ref{fig:S01pattern} shows the target and the proximity-corrected pattern used to write the nanowaveguide that hosts QD~1-3. A close-up of the grey-scale pattern taper tip is also shown. The electron dose is encoded linearly in the 256 grey-scale steps.

 \newpage
\section*{III. Power-dependent emission red-shift}

\begin{figure}[h]
\centering
\includegraphics[width=6 cm]{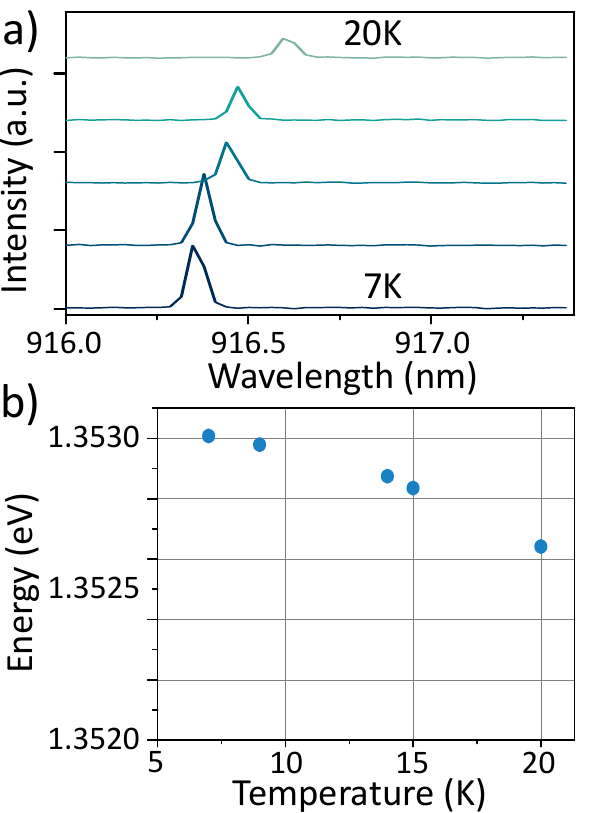}
\caption{a) Temperature series taken at temperatures of 7\,K, 9\,K, 14\,K, 15\,K and 20\,K. We added an arbitrary intensity offset in the spectra for better visualization. b) Peak energies from a) against cold-finger temperature.}
\label{fig:S1redshift}
\end{figure}

To investigate the red-shift of the $\approx916.3$\,nm emission line under p-shell excitation, we assume a linear increase $T=T_0+\eta~ P$ in temperature with excitation power $P$, starting from a base temperature of $T_0=7$~K, and fit this dependence with a Bose-Einstein phonon law~\cite{Ortner2004,ref:Davanco_BE} to the power-dependent peak positions $E_{\text{QD}}(P)$, see Fig.~3~c) in the main text.
\begin{eqnarray}
E_{\text{QD}}(P) = E_{\text{QD}}(0) - S~ E_{\text{Ph}} ~\text{coth}\bigg( \frac{E_{\text{Ph}}}{2~ (T_0+\eta~ P)~ k_{\text{B}}} \bigg)
\end{eqnarray}
Here, $S=0.5\pm0.1$ gives the phonon coupling strength, $E_{\text{Ph}}=(1.71\pm0.06)$\,meV is the phonon energy and $\eta=(6700\pm1400)$~K/W is the power-temperature-coefficient (all uncertainties are standard errors). As can be seen from Fig.~3~c) in the main text, the model represents the data very well, hinting that the temperature increases linearly with excitation power. To further investigate, we take a temperature series of the $\approx916.3$\,nm emission line, shown in Fig.~\ref{fig:S1redshift}~a). In this series, we measure the temperature within the cryostat cold finger and assume that it is equal to the actual GaAs nanowaveguide temperature, which is not directly measurable. We find that the emission energies measured in the temperature series, see Fig.~\ref{fig:S1redshift}~b), red-shift faster than predicted using the $T=T_0+\eta~ P$ law. This means that a more elaborate model is needed to fully describe the thermal behaviour of our hybrid system. During the power series experiment, the excitation wavelength remained unchanged.

 \newpage
\section*{IV. Hanbury Brown and Twiss interferometer evaluation}
The HBT data shown in Fig~3~d) in the main text is modeled with a two-sided exponential decay function, taking into account only the faster decay rate $\tau_{\text{r}}$ for a conservative estimate of $g^{(2)}(0)$:
\begin{eqnarray}
g^{(2)}(\tau) = A_0~ \text{exp}(-|\tau| / \tau_{\text{r}}) +A_{\text{Side}}~ \sum_{k\neq 0} \text{exp}(-|\tau - \tau_{\text{k}}| / \tau_{\text{r}})
\end{eqnarray}
Here, $A_0$ is the central peak area, $A_{\text{Side}}$ the side peak area, $\tau_{\text{r}}$ the radiative lifetime and $\tau_{\text{k}}$ the side peak position. This model is convolved with the detector time response of 129\,ps to obtain $g^{(2)}_{\text{Conv}}(\tau)$ for the fit. Since the coincidence data around $\tau=0$ is close to 0, we fit our model using a logarithmic Poissonian noise distribution\cite{Kirsanske2017a} $\text{ln}(\text{Poiss}(\mu,K))= -\mu +K ~\text{ln}(\mu)-\text{ln}(K!)$ and a maximum likelihood routine that minimizes
\begin{eqnarray}
-\sum_i \text{ln} \bigg[ \text{Poiss}\bigg(g^{(2)}_{\text{Conv}}(\tau_i)~,~N_i\bigg)\bigg].
\end{eqnarray}
Here, $i$ enumerates the time bins, $\tau_i$ is the time delay in bin $i$ and $N_i$ is the number of coincidences in bin $i$. The peak areas $A_0$ and $A_{\text{Side}}$ are the only free parameters in the fit, yielding $g^{(2)}(0)= A_0 / A_{\text{Side}}=0.11\pm0.04$, where the uncertainty gives the propagated 95\,\% confidence interval.

 \newpage
\section*{V. Upper bound estimate for the QD-waveguide and QD-lensed optical fiber coupling efficiency}
To estimate the maximum expected coupling efficiency between QD~3 and the fundamental TE Si$_3$N$_4$ waveguide mode, we used Finite Difference Time Domain (FDTD) simulations of electric dipoles emitting in a hybrid waveguide/mode transformer that approximated the geometry observed in SEM. In this model, the GaAs, SiO$_2$ spacer and Si$_3$N$_4$ layers had a thickness of 190\,nm, 100\,nm and 250\,nm, respectively. The straight section of the GaAs waveguide had a width of 620\,nm and length of 5\,\textmu m, and the mode transformer was tapered down to a width of 100\,nm over a length of 20\,\textmu m. The GaAs nanowaveguide and the underlying, 685\,nm width Si$_3$N$_4$ waveguide were horizontally misaligned from each other by 44\,nm. The FDTD simulations consisted of exciting the geometry with an electric dipole source located at the geometrical center of the GaAs nanowaveguide, and calculating the steady-state fields at a 916\,nm wavelength, at the edges of the computational domain.

The simulation incorporated perfectly-matched layers to emulate open domains. The total emitted power of the dipole was obtained by integrating the steady-state Poynting vector over all of the computational domain boundaries. The power carried by the various guided modes supported by the GaAs and Si$_3$N$_4$ waveguides were obtained through overlap integrals with the steady-state field at the waveguides' cross-section. Simulations were performed for  horizontally oriented dipole moments, either transversal ($x$-oriented) or longitudinal ($z$-oriented) to the GaAs nanowaveguide. Because we believe the $\approx916.3$\,nm emission line of QD~3 to be from a charged exciton, which would have circularly polarized transitions, we have also simulated the case of a rotating dipole (the $x$- and $z$-components with a $90^{\circ}$ phase between them)\cite{young_polarization_2015}. However, because a transversal dipole tends to couple more efficiently to the fundamental TE-like GaAs mode, it yields a more conservative upper-bound estimate for the QD-Si$_3$N$_4$ waveguide coupling.

\begin{figure}[h]
\centering
\includegraphics[width=14 cm]{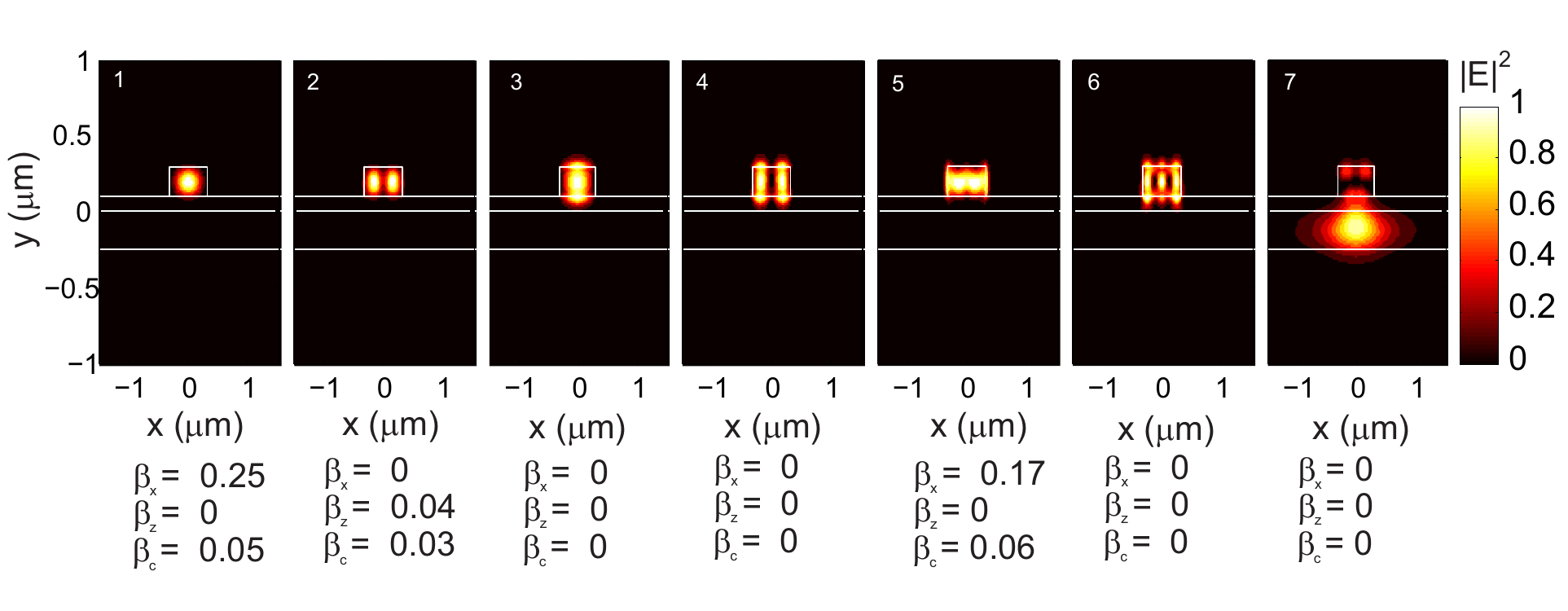}
\caption{Squared electric field profiles for bound modes of the hybrid nanowaveguide hosting QD~3. Modes 1, 2 and 5 have TE-like character, with major transversal ($x$) electric field component. Modes 3 and 4 have a TM-like character, with a major $x$ magnetic field component. The $\beta_{x,z,c}$ factors are the coupling between transversal ($x$), longitudinal ($z$) or circular ($c$) dipoles to the corresponding modes. }
\label{fig:GaAs_mode_beta}
\end{figure}

\begin{figure}[h]
\centering
\includegraphics[width=8 cm]{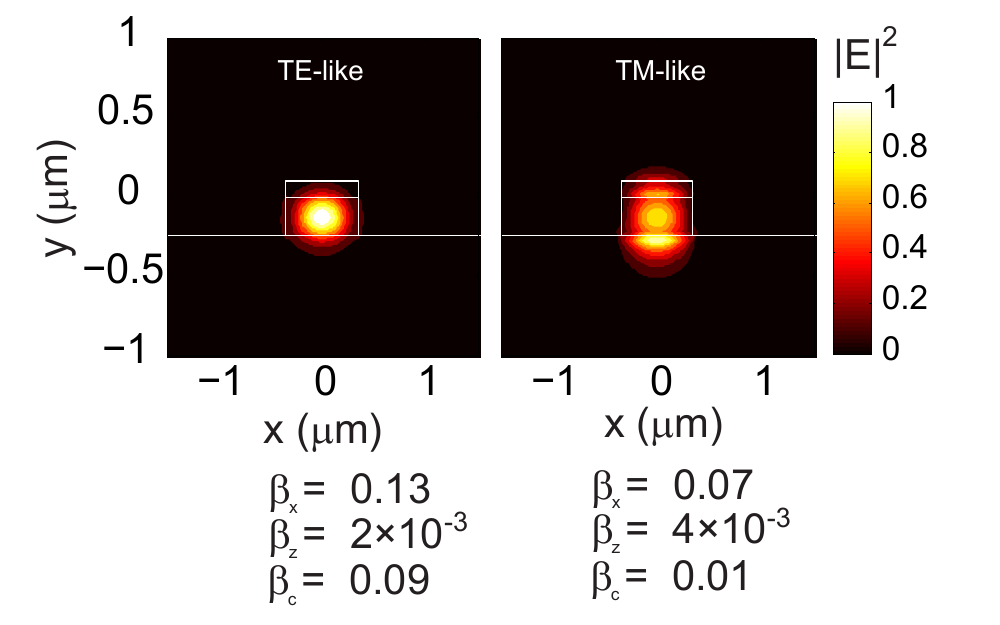}
\caption{Squared electric field profiles for bound modes of the Si$_3$N$_4$ waveguide. Modes 1 and 2 are TE-like and TM-like, respectively. The $\beta_{x,z,c}$ factors are the coupling between transversal ($x$), longitudinal ($z$) or circular ($c$) dipoles to the corresponding modes.}
\label{fig:Si3N4_mode_beta}
\end{figure}

The 620\,nm wide GaAs waveguide supports 7 guided modes at 916\,nm, as shown in Fig.~\ref{fig:GaAs_mode_beta}. The dipole emission is divided among such guided modes, as well as unguided, radiation or substrate modes. Figure~\ref{fig:GaAs_mode_beta} also shows the coupling ratios ($\beta_{x,y,z}$) for each guided mode, for the transverse ($x$), longitudinal ($z$), and rotating dipole ($c$) cases. The highest coupling ratio achievable in such multimode waveguides is of about 25$\,\%$, for the horizontal dipole into the fundamental GaAs mode, which is TE-like, and has a major $x$ electric field component. It is worth noting that about 17\,$\%$ of the dipole emission is coupled to mode 3, which is a third-order TE-like mode. Both modes 1 and 3 are converted into the fundamental Si$_3$N$_4$ waveguide TE mode, however, the conversion efficiency is highest ($\approx69\,\%$) for the fundamental one. Figure~\ref{fig:Si3N4_mode_beta} shows the fundamental TE-like and TM-like Si$_3$N$_4$ waveguide modes, after the mode transformer. The coupling efficiencies from the QD to the these two modes, for the tree dipole configurations, is also displayed. A maximum efficiency of $\approx13\,\%$ is achieved for the transverse dipole, whereas for the rotating dipole the efficiency is of $\approx5\,\%$.

We also estimated the coupling efficiency between the Si$_3$N$_4$ waveguide and the lensed optical fiber used in our experiments. The manufacturer specification for the lensed fiber was such that it should produce at $\approx2\,$\textmu m spot size at its focus, at 980\,nm, and so in our simulation, a 2\,\textmu m spot-size, horizontally polarized Gaussian beam was launched at the geometrical center of a 685\,nm wide and 250\,nm tall Si$_3$N$_4$ waveguide facet. We note that, in our fabricated devices, the Si$_3$N$_4$ waveguides unintendedly made a $\approx8^{\circ}$ angle with respect to the cleaved facet plane. This imperfection was included in our model. An overlap integral was then used to obtain, from the steady-state waveguide field, a coupling ratio of $\approx23\,\%$ into the fundamental TE-like Si$_3$N$_4$ mode.

Overall, we expect the QD-lensed fiber coupling efficiency to be of at most $0.23\times0.13\approx3\,\%$, achieveable with a transverse dipole, for a conservative estimate. We note however that for a rotating dipole, the maximum possible coupling efficiency is estimated to be $\approx1\,\%$.

 \newpage
\section*{VI. Fabry-Perot interferometer evaluation}
\begin{figure}[h]
\centering
\includegraphics[width=14.6 cm]{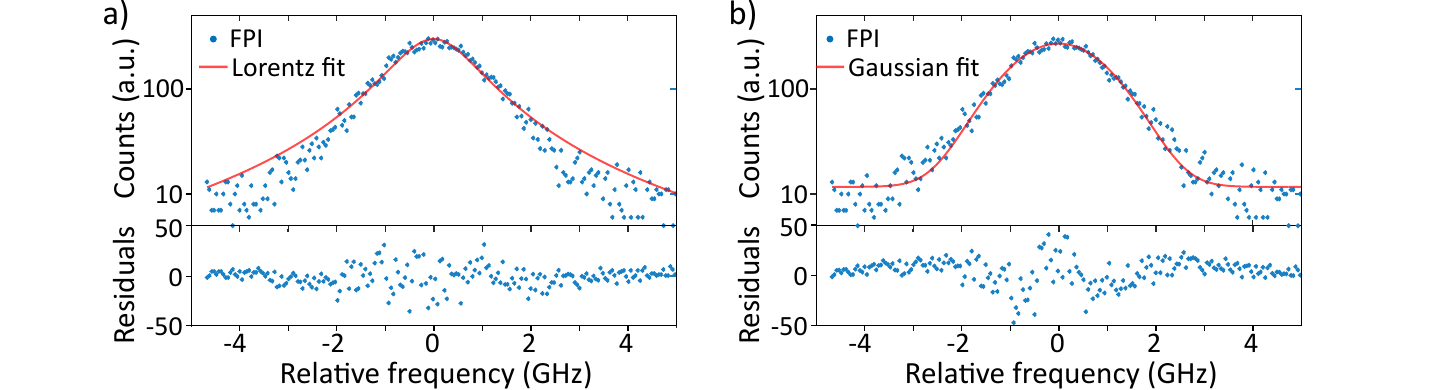}
\caption{a) FPI data (blue) and Lorentzian fit (red) with residuals. b) FPI data (blue) and Gaussian fit (red) with residuals}
\label{fig:S5FPI}
\end{figure}

For the evaluation of the Fabry-Perot-Interferometer (FPI) data, we fit the data with a Voigt~\cite{abrarov_rational_2015} (Fig.~4~a) in the main text), Lorentzian (Fig.~\ref{fig:S5FPI}~a)) and Gaussian (Fig.~\ref{fig:S5FPI}~b))  line functions. The Voigt fit gives an $R^2$ of 0.9884, slightly superior to the Lorentzian and Gaussian ($R^2=0.9766$ and $R^2=0.9867$, respectively). Inspection of the fits and residuals indicates the Voigt function gives qualitatively better results.

 \newpage
\section*{VII. Hong-Ou-Mandel interferometer}
In the Hong-Ou-Mandel (HOM) type experiment, the fiber-collected PL was first passed through a $\approx 500\,$pm free spectral range fiber-coupled grating filter followed by a long wavelength pass filter, and then guided through quarter- and half-wave plates and a polarization beam splitter (PBS) cube. This was done to bring the polarization of the PL signal into a linear state from an elliptical state that resulted in large part due to scrambling in the non-polarization-maintaining SM collection fiber, though likely also from the inherent polarization of the collected QD emission. The linearly-polarized light was then passed through a half-wave plate and coupled into a polarization-maintaining (PM) fiber PBS. The half-wave plate was aligned to the slow-axis of the fiber PBS, such that the photon stream was maximized at one output port while being suppressed at the other. Throughout the experiment, we could monitor the suppressed polarization output of the fiber PBS at an SNSPD, to verify the long term stability of our polarization filtering system, which could change if the stress on the conventional SM fibers in the setup was accidentally altered. After the fiber-coupled PBS, the QD signal was guided into an unbalanced PM fiber-coupled Mach-Zehnder Interferometer (MZI), with a $\delta \tau \approx10$\,ns arm imbalance. A variable half-wave plate inserted in the long interferometer arm allowed the (linear) polarization of photons travelling through either arm to be, at the second beamsplitter, parallel or orthogonal to each other. We matched the intensity of the photon streams in the two MZI arms by controllably loosening one fiber connection in the short arm, and measured coincidences on the two MZI outputs using the same SNSPDs as in the HBT measurement.

The raw HOM autocorrelation traces for parallel and orthogonal photons are depicted in blue in Fig.~4~b) and magnified around $\tau=0$ in Fig.~4~c) in the main text. We follow Ref.~\cite{Patel2008} to model the two traces with the functions $g^{(2)}_{\text{HOM},\parallel}(\tau)$ and $g^{(2)}_{\text{HOM},\perp}(\tau)$, where $g^{(2)}(\tau)$ accounts for the bunching around zero time delay:
\begin{eqnarray}
g^{(2)}(\tau) &=&  1- A_{1}~ \text{exp}(-|\tau| \, /\, \tau_1)+(A_1-1)~ \text{exp}(-|\tau| \, /\, \tau_2)\\
g^{(2)}_{\text{HOM},\perp}(\tau) &=& \underbrace{4 (T_1^2 + R_1^2)~R_2 T_2 g^{(2)}(\tau)}_{G_1(\tau)} +\, \underbrace{4R_1 T_1[T_2^2 g^{(2)}(\tau-\delta \tau) + R_2^2 g^{(2)}(\tau+\delta \tau)]}_{G_2(\tau)}\\
g^{(2)}_{\text{HOM},\parallel}(\tau) &=& G_1(\tau) + G_2(\tau) [1-V~ \text{exp}(-2\,|\tau |\, /\, \tau_{c,\text{HOM}})\,]
\end{eqnarray}
Here, $A_1$, $\tau_1$ and $\tau_2$ describe the bunching around $\tau=0$ in the CW autocorrelation curve $g^{(2)}(\tau)$. $R_1=0.50$ and $T_1=0.50$ as well as $R_2=0.54$ and $T_2=0.46$ are the reflection and transmission coefficients of the first and second beamsplitter in the Mach-Zehnder interferometer. From the fit of $g^{(2)}_{\text{HOM},\perp}(\tau)$, we obtain $A_1=1.64\pm0.01$, $\tau_1=(0.30\pm0.02)$\,ns and $\tau_2=(14.61\pm0.32)$\,ns, where the uncertainties are standard errors. The results of $g^{(2)}_{\text{HOM},\parallel}(\tau)$ are given in the main text.

 \newpage
\section*{VIII. Resonant and phonon-mediated excitation}
\begin{figure}[h]
\centering
\includegraphics[width=8.25 cm]{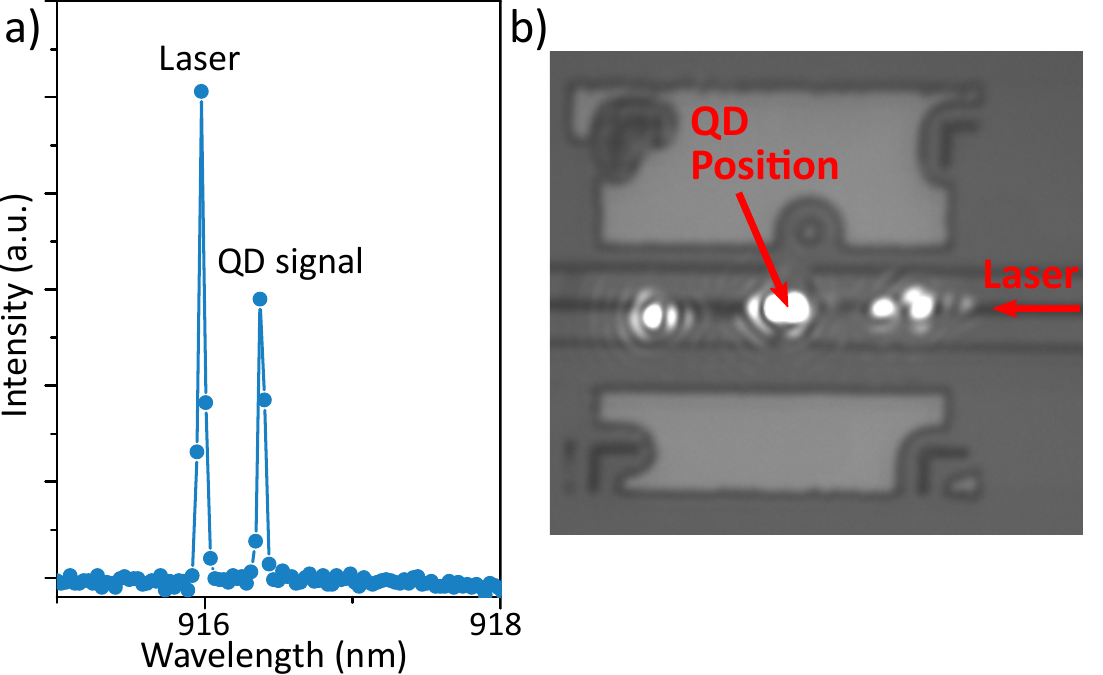}
\caption{a) \textmu PL spectrum of QD~3 showing the excitation laser and the QD signal in an LA-phonon-assisted excitation (laser excitation from top, detection from side). The laser is suppressed with an SNR of $\approx$ 1:2. b) Microscope image taken of device QD~3 with moderate white light illumination, while at the same time launching 916.3\,nm laser light into the Si$_3$N$_4$ WG from the cleaved sample edge. Large amounts of laser are scattered towards the top, as shown by bright white intensity spots. Scattering is particularly strong close to the QD position, and at the GaAs taper tips.}
\label{fig:S2phonon}
\end{figure}

The coherent excitation of QDs through resonance fluorescence is an important step towards the emission of Fourier-limited photons~\cite{Kuhlmann2015}. In resonance fluorescence, the QD signal needs to be separated from the excitation laser. Waveguide architectures naturally offer spatial separation of pump and detection position, for free-space excitation orthogonal to the wafer. Furthermore, waveguides can act as polarization filters~\cite{Schwartz2016b} allowing on-chip suppression of pulsed excitation lasers up to a signal-to-noise ratio (SNR) of 40:1~\cite{Schwartz2018}. When tuning the excitation laser wavelength close to the QD emission, one can excite the QD through a longitudinal acoustic (LA) phonon-mediated process. Assuming that the LA-phonon mediated and strictly resonant excitation are approximately equally efficient, we use this scheme to probe the QD emission to laser SNR that we can achieve in our system. Applying off-chip polarization filtering as in the HOM experiment and an excitation NA of 0.28, we measure an SNR of about 1:2 as seen in Fig.~\ref{fig:S2phonon}~a). We check the origin of the excess laser signal in our system with two methods. Firstly, we use an excitation NA of 0.1 and repeat the phonon-mediated pumping, losing one order of magnitude in SNR. Secondly, we launch resonant laser light into the Si$_3$N$_4$ waveguide at the sample facet, and measure laser light that is scattered into an NA=0.28 towards the top with a conventional CCD camera, see Fig.~\ref{fig:S2phonon}~b). Most of the light is scattered at the GaAs WG taper tips and in the center of the GaAs WG, where the QD is located. Due to time reversal symmetry, we conclude that our current GaAs-Si$_3$N$_4$-$\text{SiO}_{2}$ structure unfortunately scatters large amounts of the top excitation laser into the Si$_3$N$_4$ WG, detrimental to laser suppression. With improved waveguide designs, that avoid vertical interfaces near the QD, and a tighter beam focusing reducing the necessary pump powers, higher laser suppression enabling resonance fluorescence from preselected QDs on silicon chips may be possible in the future.

\end{document}